# A single determinant for the rate of yeast protein evolution

D. Allan Drummond[1], Alpan Raval[2,3], and Claus O. Wilke[2,1]*
[1]Program in Computation and Neural Systems
California Institute of Technology
Pasadena, CA 91125-4100, USA
[2]Keck Graduate Institute
Claremont, CA 91711, USA
[3]School of Mathematical Sciences
Claremont Graduate University
Claremont, CA 91711, USA

*Corresponding author:    Claus O. Wilke
                          Keck Graduate Institute
                          535 Watson Drive
                          Claremont, CA 91711, USA
                          Tel:  (909) 607 0139
                          Fax: (909) 607 9826
                          E-mail:  wilke@kgi.edu

Manuscript information: 41 pages, 3 figures, 3 tables

# SUMMARY

A gene's rate of sequence evolution is among the most fundamental evolutionary quantities in common use, but what determines evolutionary rates has remained unclear. Here, we show that the two most commonly used methods to disentangle the determinants of evolutionary rate, partial correlation analysis and ordinary multivariate regression, produce misleading or spurious results when applied to noisy biological data. To overcome these difficulties, we employ an alternative method, principal component regression, which is a multivariate regression of evolutionary rate against the principal components of the predictor variables. We carry out the first combined analysis of seven predictors (gene expression level, dispensability, protein abundance, codon adaptation index, gene length, number of protein-protein interactions, and the gene's centrality in the interaction network). Strikingly, our analysis reveals a single dominant component which explains 40-fold more variation in evolutionary rate than any other, suggesting that protein evolutionary rate has a single determinant among the seven predictors. The dominant component explains nearly half the variation in the rate of synonymous and protein evolution. Our results support the hypothesis that selection against the cost of translation-error-induced protein misfolding governs the rate of synonymous and protein sequence evolution in yeast.

# INTRODUCTION

A protein's evolutionary rate, commonly measured by the number of nonsynonymous substitutions per site in its encoding gene, is routinely used to characterize functional importance, detect selection (Kellis et al. 2004), create phylogenetic trees (Kurtzman and Robnett 2003), identify orthologous genes (Wall et al. 2003), and infer the time of major evolutionary events. However, what determines a protein's evolutionary rate has remained the subject of active speculation and ongoing research (Pál et al. 2001; Akashi 2003; Rocha and Danchin 2004).

Recently, studies have found significant influences on evolutionary rate from many disparate variables: proteins have been reported to evolve slower if their encoding genes have a higher expression level in mRNA molecules per cell (Pál et al. 2001), if they have a higher codon adaptation index (CAI) (Wall et al. 2005), more protein-protein interactions (higher "degree") (Fraser et al. 2002), shorter length (Marais and Duret 2001), a smaller fitness effect upon gene knockout (higher "dispensability") (Hirsh and Fraser 2001; Yang et al. 2003), or a more central role in interaction networks ("betweenness-centrality," or simply "centrality") (Hahn and Kern 2005).

Here, we first demonstrate that the analytical techniques widely used to establish independent roles for many effects, partial correlation and multiple regression, generate highly significant but entirely spurious effects given noisy data such as those available for evolutionary analyses. Then, using a technique which does not suffer from these problems, we carry out a comprehensive analysis designed to uncover the major independent correlates of evolutionary rate in the model eukaryote *Saccharomyces cerevisiae*. We determine the number of such correlates, their strength, and their relationship to the biological variables used in previous studies. Finally, we ask what these correlates reveal about the biological constraints on protein sequence evolution.

# RESULTS

**Correlation and partial correlation analysis**

We used the yeast *S. cerevisiae* to examine the determinants of evolutionary rate because it has been the subject of many previous analyses (*e.g.*, Pál et al. 2001; Fraser 2005) and has an enormous amount of available genomic, proteomic, and functional data. We first examined the raw correlation of six previously assessed biological variables (expression, CAI, length, dispensability, degree, and centrality) with protein evolutionary rate, as measured by the number of nonsynonymous substitutions per site in the underlying gene. A seventh variable, the number of protein molecules per cell ("abundance"), was also considered. Table 1 shows that all variables except centrality correlated with evolutionary rate with high significance, as previously reported.

Expression level strongly correlates with evolutionary rate, and higher-expressed genes have higher CAIs (Akashi 2001), are less dispensable (Gu et al. 2003), more abundant (Ghaemmaghami et al. 2003), and more likely to be found in protein-protein interaction experiments (Bloom and Adami 2003) than lower-expressed genes. No inverse relationships have been posited by which these variables alter expression level. Thus, it is imperative to establish whether these variables play a role independent of expression level. Following previous analyses (Pál et al. 2003; Lemos et al. 2005; Wall et al. 2005), we computed the partial correlation of our seven variables with evolutionary rate, controlling for expression level. Table 1 shows that CAI, dispensability and degree all showed reduced but highly significant partial correlations, consistent with previous studies (Hirsh and Fraser 2003; Wall et al. 2005), as did abundance.

**Partial correlations and noisy data**

What can we conclude from highly significant partial correlations? Yeast expression-level measurements from multiple groups, even two using the same commercial oligonucleotide array, correlated with coefficients of only 0.39 to 0.68 (Coghlan and Wolfe 2000), demonstrating that expression level measurements either are inaccurate and/or simply reflect the variability of gene expression across growth conditions and strains. We refer to all such variability as noise, regardless of its source. Noisy data are the rule in genome-wide molecular studies, leading us to explore what effect noise has on partial correlation analyses. As a concrete example, CAI is so tightly bound to expression level that a recent analysis used CAI as its preferred expression-level measurement (Wall et al. 2005). Might CAI's significant partial correlation only reflect our inability to control for the true (*i.e.*, evolutionarily relevant) underlying expression level? More generally, we can ask: What is the expected partial correlation of two variables, controlling for a third, when *i*) the two variables relate only through dependence on the third "master" variable, and *ii*) all measurements contain noise?

Given these conditions, we derive explicit formulas for the expected partial correlation, its statistical significance, and its behavior under various limiting cases in the Appendix. The expected partial correlation is, in general, larger than zero, because the full correlation reflects the true underlying master variable's influence, while partial correlations can only remove the portion of this influence that is visible through a noisy measurement (Box 1). We show that, surprisingly, if measurements of an underlying causal variable (*e.g.* expression level) are sufficiently noisy, highly significant partial correlations of virtually any strength between the dependent predictors can be obtained.

As a case in point, dispensability's role has been vigorously debated (Hirsh and Fraser 2003; Pál et al. 2003; Wall et al. 2005) with correlation and partial correlations acting as key analytical tools. Given a model in which expression level $X$ and noise completely determine dispensability $D$ and evolutionary rate $K$ (see Box 1), what is the observed partial correlation $r_{DK|X'}$ if we fit variables to approximately match the observed correlations between $X'$, $D$ and $K$? As a concrete example, previous reports show that, using parametric Pearson's correlations, $r_{X'K} \approx -0.6$ (Pál et al. 2001; Wall et al. 2005), $r_{DK} \approx 0.25$ (Wall et al. 2005), $r_{DX'} \approx 0.2$ (Pál et al. 2003), and $r_{DK|X'} \approx 0.24$ (Wall et al. 2005).

We can obtain roughly the reported full correlations and $r_{DK|X'} \approx 0.23 \pm 0.02$, $P \ll 10^{-9}$ with 3,000 observations if the true expression level *X* is normally distributed with mean 0.5 and standard deviation 0.25 and the observable predictors *X'*, *D*, and *K* are equal to *X* plus zero-mean normally distributed noise with standard deviations of 0.3, 0.7 and 0.1, respectively. This highly significant partial correlation is entirely spurious: in this model, expression level and random noise completely determine dispensability. Thus the observed statistical relationship between dispensability and evolutionary rate, established by correlation and partial correlation, would arise *even if no actual relationship existed* except mutual dependence on noisily measured expression level.

**Multivariate regression analysis**

Since partial correlation analysis is not applicable to the problem at hand, what other methods can we use to determine the relative influence of different predictors on the rate of evolution? One obvious choice is multivariate regression analysis, a method with the added benefit that we can look at the influence of all potential predictor variables at the same time, and can eliminate step by step those predictors that contribute the least to the regression model. Indeed, several authors have followed this route (Rocha and Danchin 2004; Agrafioti et al. 2005). Regressing dN simultaneously against the seven predictors we consider here, we find that all but centrality make a significant contribution to the regression, and that the overall $R^2 = 0.45$.

Unfortunately, ordinary multivariate regression is not appropriate to analyze the influence of the various predictors on the evolutionary rate either (Box 1). The problem is that the predictors strongly correlate with each other, while multivariate regression implicitly assumes that the predictors are statistically independent. This problem is widely discussed in the statistical literature, mostly in the context of *collinear* or *nearly collinear* predictors (Gunst and Mason 1977a,b; Mandel 1982; Næs and Martens 1988).

**Principal component regression analysis**

An alternative approach is to first identify independent sources of variation in the data, and then determine the contribution of each biological predictor to each source. The technique of principal component regression offers one way to carry out such an analysis.

In principal component regression (Mandel 1982), multiple linear predictors (*e.g.* expression level, dispensability, etc.) are scaled to zero mean and unit variance, inserted in a matrix, and rotated such that the new coordinate axes point in the directions of greatest predictor variation. The new axes define variables, called principal components, which are linear combinations of the original predictors. Subsequent linear regression of the response (*e.g.*, dN) on the rotated predictor data yields several pieces of information per principal component: the proportion of the response's variance, $R^2$, explained by the component, the significance of this $R^2$, and the fractional contribution of each original predictor to the component. Because all principal components are orthogonal and independent, the total response variation explained by the data is the sum of the component $R^2$'s. Principal component regression thus circumvents the debilitating problems of partial correlation and multivariate regression analyses (Box 1) while yielding results that are, in some ways, easier to interpret.

We carried out principal component regression on the seven predictors analyzed above. Because the determination of principal components involves only the predictors and not the response (*i.e.*, dN or dS), there is only one set of components and contributions from biological predictors. The regression analysis generates response-specific results, in particular, the proportions of variance in dN, dS, and so on, which each component explains. Table 2 shows numerical data from the analysis of dN and dS using the seven predictors of expression, CAI, abundance, length, dispensability, degree and centrality; Figures 1a and 2a show these data graphically.

Strikingly, for the rate of protein evolution, dN, one principal component explained 43% of the variance with high significance, while all other components explained less than 1% (Fig. 1). The single dominant component was almost entirely (>90%) determined by roughly equal contributions from three predictors: expression level, abundance, and CAI.

While the causes of dN's variation have remained unclear, dS is constrained by translational selection: selection for preferred codons, which correspond to abundant tRNAs and are translated faster and more accurately (Akashi 1994, 2001), makes many synonymous changes unfavorable and thus reduces dS (Hirsh et al. 2005). Figure 2 shows that the dS results mirror those using dN: the first component, which is determined almost entirely by expression, abundance and CAI, is overwhelmingly dominant (50.8% of dS variation). A second highly significant component of modest size (6.6% of dS variation) appears, but is 88.4% determined by abundance and CAI. Astonishingly, the seven biological predictors explain a cumulative 61.9% of the total variance in dS, with three predictors (expression, abundance, and CAI) contributing roughly equal amounts and accounting for 87% of the total variance explained. Because synonymous sites are thought to be under relatively weak selection, we would expect random fluctuations (noise) to contribute a large proportion of variation in dS, yet our analysis suggests that selective pressures, even those revealed using noisy data, account for almost two-thirds of the dS variation among these genes.

The size of the seven-component data set (568 genes) was severely limited by the requirement for genes having measures for all seven predictors. In particular, we used high-quality interactions measurements (Han et al. 2004) for degree and betweenness-centrality; eliminating these measurements, which apparently contribute negligible amounts to evolutionary rate, more than triples the data set size to 1,939 genes. We performed the same analysis on this expanded set and obtained similar results (Table 3, and Figures 1b and 2b).

It is common practice to interpret dS as the rate of selectively neutral divergence, and the ratio dN/dS as the deviation of protein evolutionary rate from neutral, putatively allowing detection of purifying selection or adaptive evolution. We analyzed dN/dS and found trends that were similar to those observed in dN and dS alone (Tables 2 and 3). The dominant principal component explained only half the variation in dN/dS compared to dN or dS, but the reason seems obvious in light of our results: dN and dS appear to reflect the same underlying selective force, so dividing one by the other removes much of the shared influence. In yeast, as in many other organisms, dS does not reflect neutral divergence but rather divergence constrained by translational selection for preferred codons, as previous authors have noted (Hirsh et al. 2005). These authors proposed an adjusted measure of dS, denoted dS', from which the influence of codon preference has been extracted (Hirsh et al. 2005). We thus analyzed dS' and dN/dS' (Tables 2 and 3), and found that for dS' the dominance of the first principal component was obliterated. While two components (component 1, mostly CAI, expression and abundance; and component 2, mostly dispensability) appeared to make small but possibly meaningful contributions ($R^2 > 6\%$) in the smaller seven- predictor data set, these contributions were effectively eliminated in the larger five-predictor data set ($R^2 < 3\%$), even though the major contributing predictors were still present. This sample-size dependence suggests that the contributions of components 1 and 2 are artifacts. Overall, our results are consistent with the previous claim that dS' has been purged of the influence of selection on synonymous sites (Hirsh et al. 2005). As additional support, the dN/dS' regression was nearly instinguishable from that of dN (Tables 2 and 3).

**Analysis of binary variables using analysis of covariance**

In all the above analyses, we found that protein-protein interactions and gene dispensability showed little or no apparent influence on the rate of protein evolution (dN) and synonymous-site evolution (dS), contrary to previous reports (Hirsh and Fraser 2001; Fraser et al. 2002; Fraser and Hirsh 2004; Wall et al. 2005). Perhaps these measures, as continuous predictors describing complex and poorly understood phenomena, display false precision, *i.e.*, they reflect real underlying effects but quantify them in overly precise ways that introduce noise. We reasoned that simpler measures, such as essentiality (the limiting case of dispensability, where essential genes are indispensable and non-essential genes include all those with dispensability > 0) and type of network interaction hub ("date" hubs interact with many partners individually, while "party" hubs do so simultaneously) (Han et al. 2004; Fraser 2005), subjected to a category-based analysis, might reveal relationships obscured by their noisy continuous counterparts.

Since expression level, CAI, and abundance have such an important effect on evolutionary rate, we have to carry out the category-based analysis controlling for the effect of these three predictors. We chose to perform an analysis of covariance ("ancova"). As the continuous variable we used the principal component of the three quantities expression level, CAI, and abundance, while we encoded the category (gene is or is not a party hub, is or is not a date hub, is or is not essential) as a binary variable. We found that party hubs evolve on average at 60% of the rate of genes with known interactions ($P < 10^{-5}$), date hubs at 92% of the rate of genes with known interactions (not significant), and essential genes at 84% of the speed of all genes ($P < 10^{-6}$) (Figure 3). The effect of gene essentiality is significant but small in magnitude, and date hubs show no significant rate constraint. As previously reported (Fraser 2005), party hubs do indeed experience a notable 40% reduction in evolutionary rate.

However, evolutionary rates in yeast span three orders of magnitude; interactions play at best a minor role in constraining rates.

# DISCUSSION

We have carried out the most comprehensive comparative analysis to date of potential determinants of nonsysnonymous (dN) and synonymous (dS) yeast gene evolutionary rates. We find that a single underlying component explains roughly half the variation in both dN and dS, and that this dominant component is almost entirely determined by gene expression level, protein abundance and codon bias as measured by the codon adaptation index (CAI).

The predictors we included in our analysis appear to explain roughly half the variation in dN and dS. Some other predictor(s) could explain the remaining half, but this seems quite unlikely, for a variety of reasons. First, a significant portion of evolutionary-rate variations are probably random, because the evolutionary process is inherently stochastic. Second, our $R^2$ estimates constitute a lower bound, because the $R^2$'s we find are attenuated by measurement noise, for example on microarray readings of gene expression (Coghlan and Wolfe 2000), systematic error, *e.g.* in some protein-protein interactions data (Bloom and Adami 2003), and time variation, for example in expression over the cell cycle (Cho et al. 1998). Finally, the true relationship between any of the predictors we examine and dN or dS is unlikely to be perfectly linear, and deviations from linearity reduce $R^2$.

Our results clearly suggest a single cause for most of the 1,000-fold variation in evolutionary rates among yeast genes, and the dominant component's three biological contributors suggest a cause: translational selection. We hypothesize that the number of translation events a gene experiences determines its evolutionary rate, and that expression, abundance and CAI are all roughly equally good predictors of the number of translation events. Translation is remarkably error-prone, with roughly 19% of average-length yeast proteins carrying a missense error (Drummond et al. 2005), and these errors can cause protein misfolding that imposes a well-known burden on the cell (Goldberg 2003) which scales with the number of translation events. Selection to reduce the number of error-induced misfolded proteins could constrain both gene sequence evolution, through pressure for preferred codons which confer increased translational accuracy, and protein sequence evolution, through pressure for protein sequences to tolerate translational missense errors without misfolding (Drummond et al. 2005). In this way, a single underlying cost can govern both synonymous and nonsynonymous evolutionary rates, consistent with our findings.

We used principal component regression for our analysis because, as we demonstrate, the more commonly employed techniques of partial correlation analysis and multivariate regression are inapplicable by assumption (in the latter case) and prone to produce spurious effects in the presence of noisy data (in both cases). By contrast, under principal component regression, the transformed predictors are orthogonal and uncorrelated, so that their relative contributions to the overall regression model can be evaluated independently and reliably. Moreover, we can extend this method to assess the influence of additional binary predictors, by carrying out an analysis of covariance (ancova) in which the covariable is given by the principal component that explains the majority of the response-variable variance.

How much dispensability and degree influence evolutionary rate has been a contentious issue. Regarding dispensability, the literature reflects disagreement over whether dispensability has any effect whatsoever on the rate of evolution, with partial correlation analyses playing a prominent evidentiary role (Hirsh and Fraser 2003; Pál et al. 2003; Wall et al. 2005). Our analysis, which avoids problematic partial correlations but uses the same data as in previous analyses that appeared to confirm a significant role for dispensability (Wall et al. 2005), is quite clear: dispensability neither constitutes an independent source of variation in dN nor contributes meaningfully to the dominant component that does influence dN. In the case of degree, the disagreement has pivoted on whether experimental surveys tend to detect interactions more often in highly expressed proteins (Bloom and Adami 2003, 2004; Fraser and Hirsh 2004), leading to a true but biologically irrelevant degree–dN relationship. Our analysis shows that degree does not contribute independently, but makes a small, significant contribution to the variable dominated by expression, abundance and CAI, as expected under the expression-bias hypothesis and inconsistent with a true constraint from the number of interactions. In short, our results suggest neither degree nor dispensability make much difference in dN, and point out precisely why previous authors have been led to the opposite conclusion.

By contrast, our ancova offers support for the observation that proteins that interact with multiple partners simultaneously, so-called party hubs, evolve slower (Fraser 2005), even after accounting for the translational effect we identify here. Party hubs epitomize the intuition behind the interactions hypothesis: while interactions presumably constrain residues, it appears that in order to slow a protein's overall rate of evolution, interactions must involve a significant proportion of the protein's residues, as expected for party hubs. Date hubs, which would include proteins that interact with many partners all at the same site serially, appeared significantly slower-evolving in the previous analysis (Fraser 2005), but our results suggest this finding reflected a failure to properly control for expression-linked effects (again, partial correlation was used). [Note that ancova enjoys a crucial and useful advantage over partial correlation aside from its greater reliability, namely the ability to control for multiple intercorrelated variables (*e.g.* expression and CAI) simultaneously.]

The rates dN and dS are routinely used to carry out analyses on selection, often under the assumption that dN/dS > 1 indicates adaptive protein evolution and dN/dS < 1 indicates purifying selection, and generally with the intent of quantifying functional pressures. Our results suggest that both evolutionary rates are determined by translational selection and are therefore likely poor predictors of functional selection, because translational selection by definition operates before a protein becomes functional. In yeast, dS does not measure neutral divergence, and thus, in the absence of a quantitative description of the relative strengths of selection on nonsynonymous and synonymous sites, the measure dN/dS is meaningless. Recently, a method for correcting dS for synonymous-site selection was proposed (Hirsh et al. 2005), and we found that the adjusted measure of neutral divergence, dS', indeed appears free of the influence of the dominant variable we link to translational selection. Our dominant variable shows virtually identical predictive power for dN and dN/dS', indicating that dividing out the neutral divergence, which for example might be due to variable mutation rates, makes little difference when analyzing the rate of protein evolution. Our results suggest that using overall gene evolutionary rates to characterize functional selection is unwise; examining dN and dS at particular sites remains a powerful and important tool in evolutionary analyses of genes.

We have found that yeast coding sequences accumulate substitutions according to a surprisingly simple formula: more predicted translation events means slower evolution. In recent years, evidence has accumulated that translation-linked variables, in particular expression levels, govern the evolutionary rate of proteins across all life, from bacteria (Rocha and Danchin 2004) to fungi (Pál et al. 2001), plants (Wright et al. 2004) and animals (Duret and Mouchiroud 2000) including humans (Subramanian and Kumar 2004), but translational selection has only recently been proposed as an explanation for this puzzling trend (Drummond et al. 2005). Our results suggest that translational selection dominates the rate of protein evolution, and by extension suggest that translational selection operates across the tree of life, from prokaryotes to humans. Future work must illuminate the precise biophysical effects that constrain molecular evolution, but we have shown that, at least in yeast, the answers may be found in translation.

## MATERIALS AND METHODS

**Genomic data**: We obtained codon adaptation indices and high-quality evolutionary rates (nonsynonymous substitutions per site dN, synonymous substitutions per site dS, adjusted synonymous substitutions dS' (Hirsh et al. 2005), and ratios dN/dS and dN/dS') from four-way yeast species alignments for 3,036 *S. cerevisiae* genes (Wall et al. 2005, supporting information, Table 4). Deletion-strain growth-rate data were downloaded from http://chemogenomics.stanford.edu/supplements/01yfh/files/orfgenedata.txt; the average growth rates of the homozygous deletion strains were used as dispensability measurements in our analysis. The FYI yeast protein interaction data set (Han et al. 2004) provided interaction network hub types for 199 genes and the number of interactions for 1,379 yeast genes. The latter data set was used to compute betweenness-centrality values, which quantify the frequency with which a network node lies on the shortest path between other nodes, as described (Hahn and Kern 2005).

**Statistical analysis**: We used R (Ihaka and Gentleman 1996) for statistical analyses and plotting. The package 'pls' was used to perform principal components regression. We log-transformed all variables besides dispensability. We decided whether or not to log-transform a variable based on whether log-transformation led to a higher $R^2$. For those variables that contained zeros, we added a small constant before the log-transformation, as suggested (Wall et al. 2005). This constant was 0.001 for dN, dS' and dN/dS', and $10^{-7}$ for betweenness centrality. We scaled the predictor variables to zero mean and unit variance before carrying out the principal component analysis. In all regression analyses (both against the original predictors and against the principal components), we determined statistical significance levels by starting with the full model and successively dropping the least-significant predictor until only significant predictors ($P < 0.01$) remained.

**Supplementary material**: All data and scripts used to perform our analyses are available as supplementary material.

**Box 1**

**Comparing partial correlation, multivariate regression and principal component regression**

How do the three analytical techniques considered here fare given a case where only one variable determines evolutionary rate? For each technique, what would we conclude about the number and strength of the rate determinants?

Consider a simple model in which a variable *X* (*e.g.*, expression level) determines two other variables, a putative determinant *D* (*e.g.*, dispensability) and a response *K* (evolutionary rate), so that $D = X + \varepsilon_D$ and $K = X + \varepsilon_K$, where $\varepsilon_D$ and $\varepsilon_K$ are noise terms with mean 0 and variances $\sigma_D^2$ and $\sigma_K^2$. Further assume that we cannot measure *X* but only a noisy correlate, $X' = X + \varepsilon_{X'}$. In this model, *X* is responsible for all the correlation between *D* and *K*. We let *X* be normally distributed with mean 0.5 and standard deviation 0.25 (so that *X* values span the unit interval) with the observable predictors *X'*, *D*, and *K* equal to *X* plus zero-mean normally distributed noise with standard deviations of 0.3. We ran each analysis 100 times with 3,000 measurements each.

Partial correlation analysis suggests that both *D* and *X'* contribute to the rate *K* independently and with equal strength:

| Partial correlation with *K* | *P*-value |
|---|---|
| $r_{DK|X'} = 0.296 \pm 0.03$ | $\ll 10^{-9}$ |
| $r_{X'K|D} = 0.291 \pm 0.02$ | $\ll 10^{-9}$ |

Multivariate regression similarly suggests that both *D* and *X'* independently influence the rate *K*:

| Predictor | % variance in $K$ explained ($R^2$) | $P$-value |
|---|---|---|
| $X'$ | $16.9 \pm 2$ | $\ll 10^{-9}$ |
| $D$ | $17.3 \pm 2$ | $\ll 10^{-9}$ |

Principal component regression, however, properly identifies only one component which contributes significantly to the rate $K$.

| Component | % variance in $K$ explained ($R^2$) | $P$-value |
|---|---|---|
| 1 | 21.3 | $\ll 10^{-9}$ |
| 2 | 0 | 0.7 |

We may proceed with the confidence that we have properly identified the number and strength of the underlying determinants of $K$. The identity of the dominant component is not given to us (in this case we know it is $X$, the true expression level), and other analyses must be used.

Our toy model underscores a key observation: In the presence of noisy and correlated data, nonzero partial correlations and $R^2$ values from multivariate regression, even those with very high statistical significance, must not be taken as evidence for independent effects, contrary to previous studies (Lemos et al. 2005; Wall et al. 2005).

# APPENDIX

**Spurious partial correlations from noisy data**

Consider a model in which a variable $X$ determines two other variables $D$ and $K$ in a linear fashion. Then we can write, without loss of generality,

$$D = X + e_D, \quad K = X + e_K, \tag{1}$$

where $e_D$ and $e_K$ are noise terms with mean 0 and variances $\sigma_D^2$ and $\sigma_K^2$, respectively. We assume that these noise terms are each independent of $X$ and of each other. In the following analysis, we also assume for convenience that $X$ has variance 1; results for the case when $X$ has an arbitrary variance $\sigma^2$ follow by dividing all other variances in the problem by $\sigma^2$ in the equations below.

The partial correlation between $D$ and $K$ given $X$ is defined as

$$r_{DK|X} = \frac{r_{DK} - r_{DX} r_{KX}}{\sqrt{(1 - r_{DX}^2)(1 - r_{KX}^2)}}, \tag{2}$$

where $r_{ij}$ is the standard Pearson correlation between variables $i$ and $j$. Given the model (1), it is intuitively obvious that $r_{DK|X}$ should be 0, because $e_D$ and $e_K$ are independent noise sources. Indeed, we find in this case

$$r_{DK} \equiv \frac{E(DK) - E(D)E(K)}{\sigma_D \sigma_K} = [(1+\sigma_D^2)(1+\sigma_K^2)]^{-1/2},$$

$$r_{DX} = (1+\sigma_D^2)^{-1/2}, \quad r_{KX} = (1+\sigma_K^2)^{-1/2}, \tag{3}$$

giving $r_{DK|X} = 0$. In the above, $E()$ denotes expected value.

We now consider the question of whether it is possible that we observe a spurious partial correlation between $D$ and $K$ if we are given a somewhat noisy version of $X$ instead of $X$ itself. Thus we introduce the new variable $X' = X + e_{X'}$, where the noise $e_{X'}$ is assumed to have mean 0 and variance $\sigma^2_{X'}$, and now compute the partial correlation $r_{DK|X'}$ since this is the actual partial correlation we would observe if we were given noisy samples of $X$. We find that $r_{DK}$ remains the same as in Eq. (3), while $r_{DX'}$ and $r_{KX'}$ are now given by

$$r_{DX'} = \left[(1+\sigma_D^2)(1+\sigma_{X'}^2)\right]^{-1/2}, \quad r_{KX'} = \left[(1+\sigma_K^2)(1+\sigma_{X'}^2)\right]^{-1/2}, \tag{4}$$

and the new partial correlation is

$$r_{DK|X'} = \frac{\sigma_{X'}^2}{\sqrt{(\sigma_D^2 + \sigma_{X'}^2 + \sigma_D^2 \sigma_{X'}^2)(\sigma_K^2 + \sigma_{X'}^2 + \sigma_K^2 \sigma_{X'}^2)}}. \tag{5}$$

Thus the presence of noise in the samples of $X$, characterized by the variance $\sigma_{X'}^2$, leads to a non-zero spurious partial correlation between $D$ and $K$. Since Eq. (5) is quadratic in $\sigma_{X'}^2$, we can also write it in terms of the amount of noise on $X$ that would result in a given spurious partial correlation:

$$\sigma_{X'}^2 = \frac{\sigma_K^2(1+\sigma_D^2) + \sigma_D^2(1+\sigma_K^2) + \sqrt{(\sigma_K^2 - \sigma_D^2)^2 + 4r_{DK|X'}^{-2}\sigma_K^2\sigma_D^2}}{2\left(r_{DK|X'}^{-2} - (1+\sigma_D^2)(1+\sigma_K^2)\right)} \tag{6}$$

We may alternatively specify a significance level as a P-value $P$ for the desired spurious correlation and the number of data points $n$ and ask what noise level in *X* is required to achieve the given significance. Based on asymptotic (large *n*) results, the P-value for testing significance of the partial correlation is given (Wall et al. 2005) as

$$P = 2[1 - \Phi(|t|)], t = r\sqrt{\frac{n-3}{1-r^2}}, \quad (7)$$

where $\Phi(x)$ denotes the standard normal cumulative distribution function and *r* is a partial correlation. Solving for the partial correlation in terms of *P* and *n*, we obtain

$$r^{-2} = 1 + \frac{n-3}{z_{1-P/2}^2}, \quad (8)$$

where $z_\gamma$ is the 100×γth percentile of the standard normal distribution. Given a P-value *P* and the number of data points *n*, we may therefore use Eq. (8) to find the corresponding partial correlation *r* and then substitute this partial correlation in Eq. (6) to find the noise level in *X* that would produce that partial correlation to the desired significance.

There are a number of important consequences and special cases of Eqs. (5) and (6):

**(i)** For fixed $\sigma_D$ and $\sigma_K$, $r_{DK|X'}$ is maximized when $\sigma_{X'}^2 \to \infty$. This maximum achievable partial correlation is given by

$$r_{DK|X'} \to \left[(1+\sigma_D^2)(1+\sigma_K^2)\right]^{-1/2}. \quad (9)$$

**(ii)** As expected, $r_{DK|X'}$ increases as the noise level on X increases. Perhaps less intuitive is the fact that the partial correlation also increases as the noise levels on D and K become smaller. In the limit that D and K are perfectly clean ($\sigma_{D,K} \to 0$), we find a perfect but spurious partial correlation of 1 with an arbitrary, non-zero $\sigma_{X'}$. Thus, if we are given noisy samples of X, we would be falsely led to conclude that D and K are well-correlated even if we control for X.

**(iii)** When $\sigma_K$ and $\sigma_D$ are comparable, i.e., $\sigma_K \cong \sigma_D = \sigma_c$, Eq. (6) simplifies to yield

$$\sigma_{X'}^2 = \frac{\sigma_c^2}{r_{DK|X'}^{-1} - (1+\sigma_c^2)}. \tag{10}$$

With $\sigma_c = 1$ (almost a worst-case scenario: in most realistic situations, $\sigma_c$ will be much smaller than the variation in X and we would then need a much smaller noise level in X in order to obtain a significant $r_{DK|X'}$) we find, using Eq. (8) with P=0.05 (95% significance) and Eq. (10), that $\sigma_{X'}^2 = 0.11$ for n=500 and $\sigma_{X'}^2 = 0.07$ for n=1000. Thus, for 1000 data points, and assuming a large noise on D and K, we only need a modest (~7% of the variance in X) amount of noise to achieve a 95% significant partial correlation. This noise level would, of course, be much lower if D and K were less noisy.

**(iv)** When n is large, Eq. (8) implies that a small partial correlation is needed to achieve significance. We may therefore assume that the $r^{-2}$ terms dominate in Eq. (6), which may then be combined with Eq. (8) to yield, for large $n$,

$$\sigma_{X'}^2 \cong \frac{\sigma_K \sigma_D z_{1-P/2}}{\sqrt{n}}, \qquad (11)$$

which directly expresses the noise level $\sigma_{X'}$ required to achieve significant partial correlation in terms of $n$ and $P$. This shows that for fixed $\sigma_D$, $\sigma_K$, and $P$ the amount of noise on $X$ that is required to achieve significance decreases as $1/\sqrt{n}$ for large $n$.

**Partial correlations in terms of measurable quantities.** Since all variances in Eqs. (5) and (6) are really ratios with respect to the true variance of $X$, it may appear that we need to know the true variance of $X$, an unmeasurable quantity, in order to find $r_{DK|X'}$. This is, however, not the case. Suppose we make two measurements of a variable $Y = X + e_Y$ (here, $Y$ could stand for any of $X'$, $D$, or $K$), where $e_Y$ is a noise source with mean 0 and variance $\sigma_Y^2$. These two measurements, say, $Y_1$ and $Y_2$, can be expressed as $Y_1 = X + e_1$, $Y_2 = X + e_2$, where $e_1$ and $e_2$ are i.i.d. noise sources with the same distribution as $e_Y$. The Pearson correlation between the two measurements $Y_1$ and $Y_2$ is then given by

$$r_Y \equiv \frac{E(Y_1 Y_2) - E(Y_1) E(Y_2)}{\sqrt{\mathrm{Var}(Y_1) \mathrm{Var}(Y_2)}} = (1 + \sigma_Y^2)^{-1}, \qquad (12)$$

where we have assumed, as before, that the variance of *X* itself is set to 1. We can therefore use Eq. (12) to express the variances of *X'*, *D* and *K* in Eqs. (5) and (6) in terms of the two-measurement correlations $r_{X'}$, $r_D$, and $r_K$, respectively, as

$$\sigma^2_{X',D,K} = r^{-1}_{X',D,K} - 1. \tag{13}$$

Such a substitution ensures that the true variance of *X* does not appear in Eqs. (5) and (9) and that these equations are expressed directly in terms of measurable quantities. In particular, Eq. (9) takes the particularly simple form $r_{DK|X'} \to \sqrt{r_D r_K}$ as $\sigma^2_{X'} \to \infty$ (or $r_{X'} \to 0$), and Eq. (10) becomes, with $r_c = r_D$ or $r_c = r_K$:

$$r_{X'} = \frac{1 - r_{DK|X'} r_c^{-1}}{1 - r_{DK|X'}}. \tag{14}$$

## ACKNOWLEDGMENTS

We are grateful to Frances H. Arnold for insightful comments on the manuscript.  This work was supported by NIH National Research Service Award 5 T32 MH19138 (to D.A.D.)

## AUTHOR CONTRIBUTIONS

DAD and COW designed the study and analyzed the data. AR derived the formulas for partial correlations under noisy data and calculated the betweenness-centrality values. All authors contributed to the writing of the manuscript.

# TABLES

**Table 1**: Partial correlation analysis of seven putative determinants of evolutionary rate.

| Variable X | Correlation $r_{X,dN}$ | Partial correlation $r_{X,dN|expr.}$ |
|---|---|---|
| Gene expression | −0.537*** | 0 |
| Codon adaptation index (CAI) | −0.565*** | −0.338*** |
| Protein abundance | −0.478*** | −0.232*** |
| Gene length | 0.136*** | 0.010 |
| Gene dispensability | 0.265*** | 0.183*** |
| Degree (# of protein-protein interactions) | −0.246*** | −0.127* |
| Protein centrality (frequency on node-node shortest paths) | −0.098 | −0.082 |

Significance codes: *, $P < 10^{-3}$; **, $P < 10^{-6}$; ***, $P < 10^{-9}$.

**Table 2**: Results of principal component regression analysis on seven predictors and five measures of evolutionary rate for 568 *S. cerevisiae* genes

| % variance explained in: | Principal components | | | | | | | |
|---|---|---|---|---|---|---|---|---|
| | 1 | 2 | 3 | 4 | 5 | 6 | 7 | All |
| dN | 42.76*** | 0.05 | 0.50 | 0.19 | 0.14 | 0.47 | 0.48 | 44.60*** |
| dS | 50.77*** | 2.13** | 0.88* | 0.08 | 6.55*** | 0.37 | 1.14* | 61.92*** |
| dN/dS | 24.82*** | 0.05 | 0.67 | 0.82 | 0.00 | 0.00 | 0.05 | 26.42*** |
| dS′ | 6.70*** | 0.19 | 7.31*** | 0.26 | 0.06 | 0.14 | 1.25 | 15.92*** |
| dN/dS′ | 42.34*** | 0.09 | 0.13 | 0.28 | 0.13 | 0.40 | 0.70 | 44.07*** |
| % contributions: | | | | | | | | |
| Expression | **32.8** | 1.2 | 1.5 | 0.1 | 1.2 | 11.2 | **52.1** | |
| CAI | **28.3** | 3.1 | 8.4 | 0.9 | 2.7 | 17.6 | **39.0** | |
| Abundance | **29.2** | 2.0 | 1.6 | 0.3 | 15.4 | **51.4** | 0.1 | |
| Length | 2.0 | 1.1 | **86.4** | 0.0 | 2.1 | 0.3 | 8.2 | |
| Dispensability | 1.8 | 13.0 | 0.0 | **84.0** | 0.0 | 0.9 | 0.3 | |
| Degree | 5.0 | **36.7** | 1.9 | 6.2 | **38.9** | 10.9 | 0.4 | |
| Centrality | 0.9 | **42.9** | 0.3 | 8.5 | **39.6** | 7.7 | 0.0 | |

Significance codes: *, $P < 10^{-3}$; **, $P < 10^{-6}$; ***, $P < 10^{-9}$. Bold indicates that the indicated predictor contributes at least 20% to the indicated component.

**Table 3**: Results of principal component regression analysis on five predictors and five measures of evolutionary rate for 1,939 *S. cerevisiae* genes

|  | Principal components | | | | | |
|---|---|---|---|---|---|---|
| % variance explained in: | 1 | 2 | 3 | 4 | 5 | All |
| dN | 36.94*** | 0.05 | 0.03 | 0.22 | 0.60*** | 37.85*** |
| dS | 39.33*** | 0.73** | 0.09 | 1.93*** | 1.92*** | 44.01*** |
| dN/dS | 22.39*** | 0.28 | 0.21 | 0.00 | 0.21 | 23.10*** |
| dS′ | 1.26** | 2.52*** | 2.58*** | 0.00 | 1.54** | 7.91*** |
| dN/dS′ | 37.61*** | 0.28 | 0.00 | 0.14 | 1.16** | 39.20*** |
|  | | | | | | |
| % contributions: | | | | | | |
| Expression | **33.2** | 1.7 | 0.1 | **24.2** | **40.8** | |
| CAI | **31.4** | 1.0 | 9.4 | 9.0 | **49.2** | |
| Abundance | **31.3** | 0.6 | 0.4 | **65.8** | 1.9 | |
| Length | 2.0 | **61.0** | **29.6** | 0.4 | 7.0 | |
| Dispensability | 2.1 | **35.7** | **60.5** | 0.6 | 1.1 | |

Significance codes: *, $P < 10^{-3}$; **, $P < 10^{-6}$; ***, $P < 10^{-9}$. Bold indicates that the indicated predictor contributes at least 20% to the indicated component.

# FIGURE LEGENDS

**Figure 1**: Principal components regression on the rate of protein evolution (dN) in 568 yeast genes reveals a single dominant underlying component. **a**, Of the seven principal components, only one (starred) explained a statistically significant proportion of the variation in dN. This component explained 43% of the variance, while no other component explained more than 1%. Expression level, codon adaptation index, and protein abundance determined most of this dominant component (labeled), while the remaining predictors (in order from top to bottom: length, dispensability, degree, centrality) determined < 10% of the component's $R^2$. See Table 2 for numerical data. **b**, A larger data set (1,939 genes) excluding protein-protein interaction predictors showed the same patterns as in **a**.

**Figure 2**: Principal components regression on the rate of synonymous-site evolution (dS) in 568 yeast genes reveals a single dominant underlying component. **a**, Seven predictor variables (see text) yielded seven principal components, of which six (starred) explained a statistically significant proportion of the variation in dS. The dominant component explained 51% of the variance, while no other component explained more than 7%. See Figure 1 caption for the breakdown of predictor contributions.  **b**, A larger data set (1,939 genes) excluding protein-protein interaction predictors showed the same patterns as in **a**.

**Figure 3**: Binary analyses of the influence of hub type and essentiality on dN reveal subtle relationships masked by continuous analyses.  **a**, Party hubs (dark points, solid line), which interact with many partners at once, evolve at 60% of the average rate for all genes with measured interactions (light points, dashed line).  **b**, Date hubs (dark points, solid line), which interact with many partners sequentially, evolve at 92% of the rate (not significant) for genes with measured interactions (light points, dashed line).  **c**, Essential genes (dark points) evolve at 84% of the genome average rate (light points, dashed line).

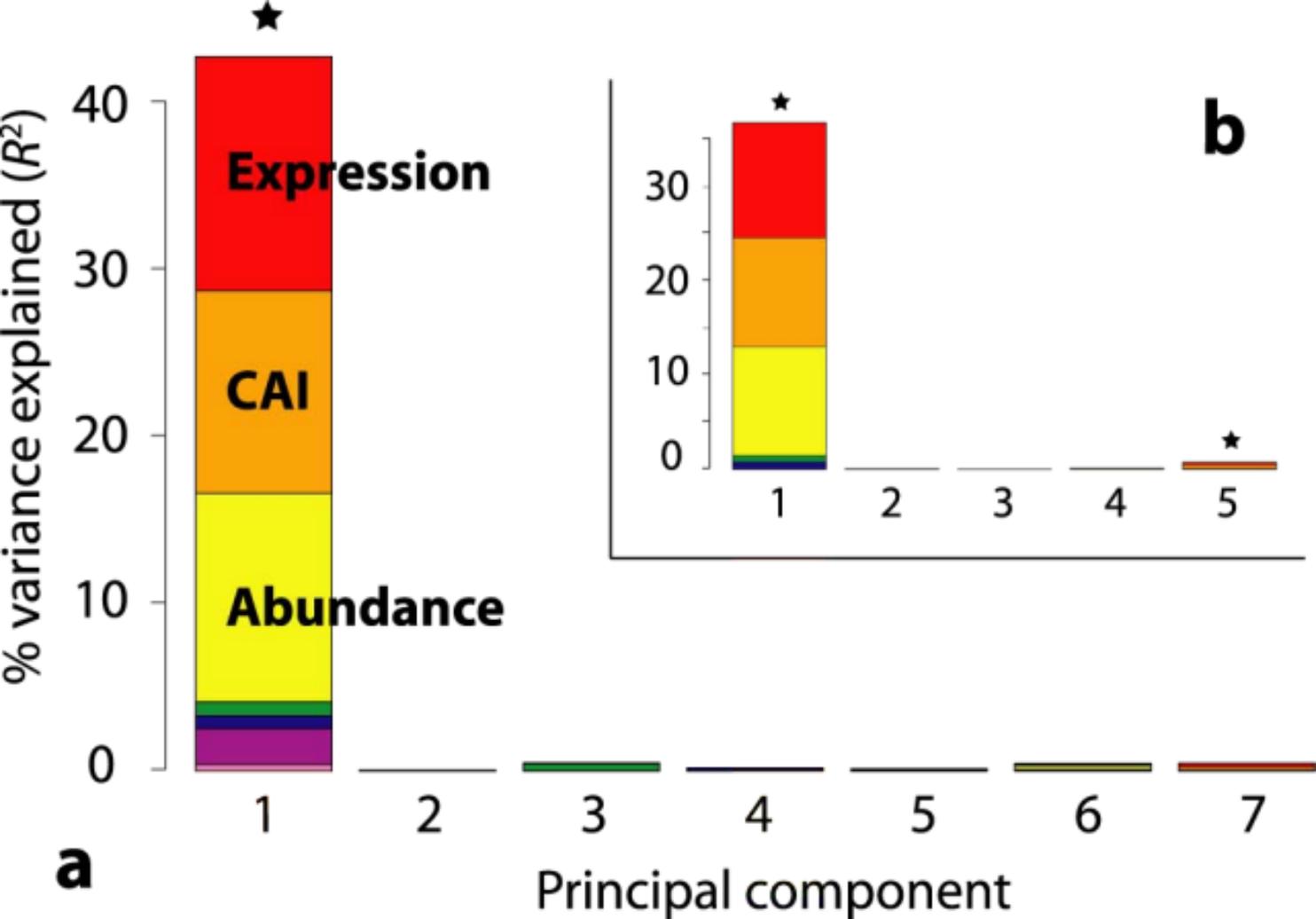

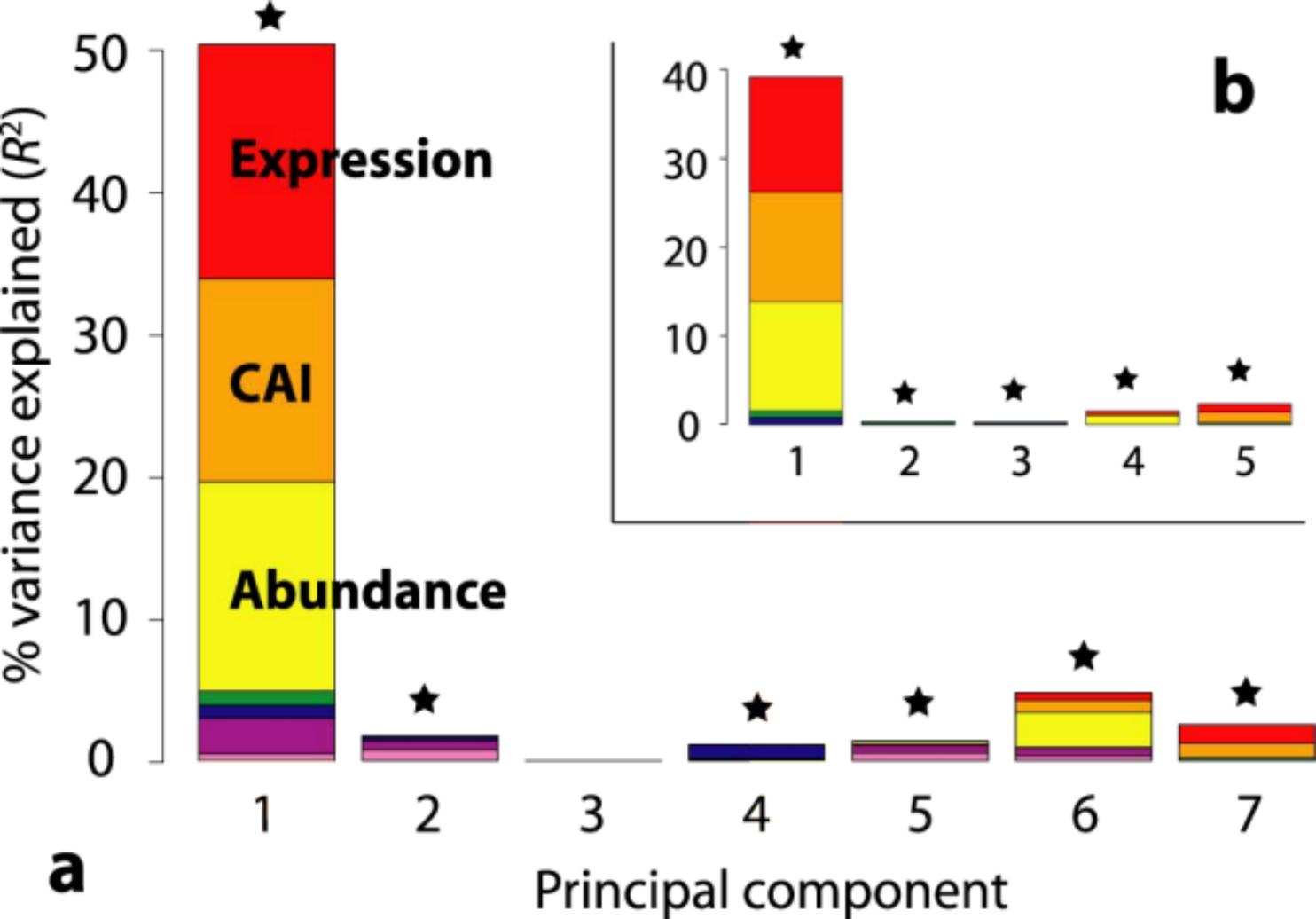

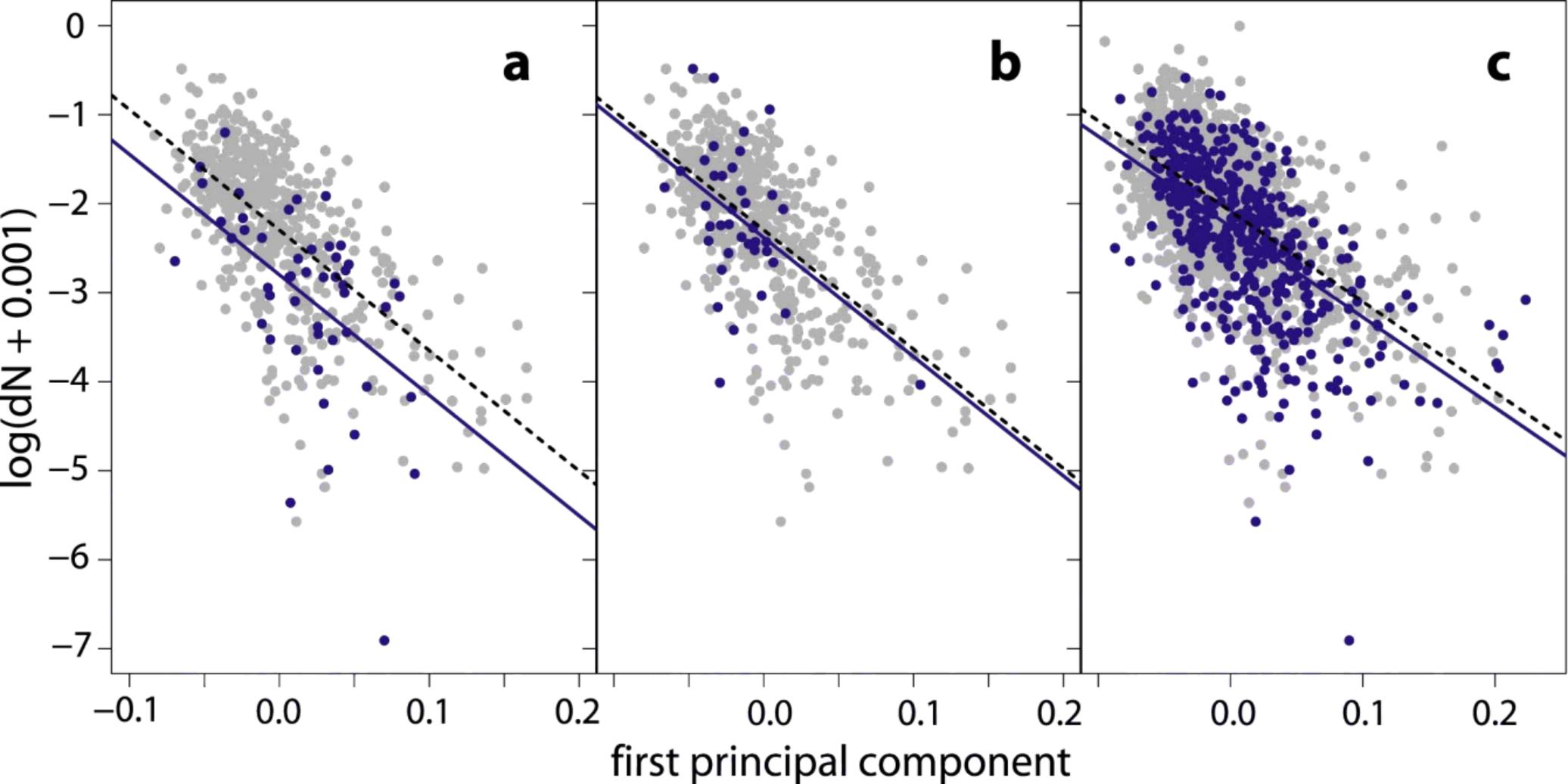